\documentstyle[emulapj,epsfig]{article}

\font\tenbg=cmmib10 at 10pt
\def \rvecphi{{\hbox{\tenbg\char'036}}}

\begin{document}

\title{Rossby Wave Instability  of
Thin Accretion Disks -- II. Detailed 
Linear Theory}

\author{H. Li\altaffilmark{1},  J.M.
Finn\altaffilmark{2},  R.V.E.
Lovelace\altaffilmark{1,3} and S.A.
Colgate\altaffilmark{1}}

\altaffiltext{1}{Theoretical
Astrophysics, T-6, MS B288, Los
Alamos National Laboratory, Los
Alamos, NM 87545. hli@lanl.gov;
colgate@lanl.gov}
\altaffiltext{2}{T-15, Los Alamos
National Laboratory,  Los Alamos, NM
87545}
\altaffiltext{3}{Department of
Astronomy, Cornell University,
Ithaca, NY 14853; rvl1@cornell.edu}

\begin{abstract}

    In earlier work we
identified a  global, non-axisymmetric
instability associated with the
presence of an extreme
in the radial profile of the key
function
${\cal L}(r) \equiv (\Sigma
\Omega/\kappa^2) S^{2/\Gamma}$ in a
thin, inviscid, nonmagnetized
accretion disk.
   Here, $\Sigma(r)$ is
the surface mass density of the disk,
$\Omega(r)$ the angular rotation
rate, $S(r)$ the specific entropy,
$\Gamma$ the adiabatic index, and
$\kappa(r)$ the radial epicyclic
frequency.
   The dispersion relation of
the instability was shown to be
similar to that of Rossby waves in
planetary atmospheres.
    In this paper,
we present the detailed linear theory
of this Rossby wave instability  and
show that it  exists for a wider range
of  conditions,
specifically, for the case where
there is a ``jump'' over some range
of $r$ in
$\Sigma(r)$ or in the pressure $P(r)$.
     We elucidate the physical
mechanism of this instability and its
dependence on various parameters,
including the magnitude of
the ``bump'' or ``jump,''
the azimuthal
mode number, and the sound
speed in the disk.
   We find large
parameter range where the
disk is stable to axisymmetric
perturbations, but
unstable to the non-axisymmetric
Rossby waves.
    We find
that growth rates of the Rossby wave
instability  can be  high, $\sim 0.2
\Omega_{\rm K}$ for relative small
``jumps'' or ``bumps''.
     We discuss
possible conditions which can lead to
this instability and the consequences
of the instability.
\end{abstract}

\section{Introduction}

   The central problem
of accretion
disk theory is understanding
the mechanism of
angular momentum
transport.
    The angular momentum must
flow  outward in order that
matter  accrete onto the central
gravitating object.
  Earlier work suggested hydrodynamic
turbulence as the mechanism of
enhanced turbulent viscosity $\nu_t =
\alpha c_s h$, with $\alpha \leq 1$ a
dimensionless constant, $c_s$ the
sound speed, and $h$ the half-thickness
of the disk
(Shakura \& Sunyaev 1973).
   However, the  physical origin
and level of the turbulence is not
established (see
\cite{paplin95} for a review).
     Recently, the magneto-rotational
instability (\cite{vel59};
\cite{chan60}; see review by
\cite{bh98}) has been studied in $2D$
and $3D$  MHD simulations and shown
to give a Maxwell  stress sufficient
to give significant outward transport
of angular momentum
(\cite{bh98}); that is,  a statistically
averaged, effective
Shakura-Sunyaev alpha parameter is
found to be $\alpha \sim 0.01$
(Brandenburg et al. 1995).
   However, significant questions
remain.
   The simulation
studies are local in that a
shearing patch or box of the actual
disk of size $\ll r$ is treated; and
the boundary conditions on the top
of the box have so far been
unphysical.
    Furthermore, disks  in
some systems (e.g., protostellar
systems) are predicted
to have very small
conductivity so that  coupling
between matter and
magnetic field is negligible.

\begin{figure*}[t]
\centering
\epsfig{file=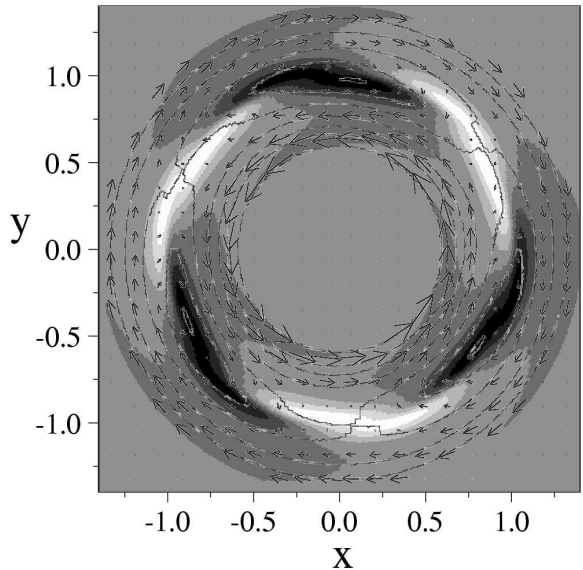,width=3in,height=3in}
\caption{
Illustration of the
nature of the Rossby wave
instability for $m=3$ for
the case of a homentropic
Gaussian bump centered
at a radius $r_0=1$.
  The gray-scale component
of the figure represents the
surface density perturbation
$\delta \Sigma$ with the darkest
region  the highest
density.
   The contour lines and the
arrows show the
velocity field as
seen in a reference frame
rotating at the angular rate
of the disk
$\Omega(r_0)$.
   This velocity
field is
$\Delta {\bf v}=\delta {\bf v}
+\hat{\rvecphi~}[r\Omega(r)-r_0\Omega(r_0)]$,
where $\delta {\bf v}$ is the
usual Eulerian velocity
perturbation.
   The figure shows that the
perturbation of the disk involves
three {\it anticyclones} of closed
streamlines or  ``islands'' which approximately
coincide with the regions of enhanced
densities $\delta \Sigma$.
   The radial width of the ``islands''
has been exaggerated to make them
visible.
\label{fig:rossby}}
\end{figure*}

   Interesting fundamental questions
still exist concerning pure
hydrodynamic processes in accretion
disks.
     One well-known hydrodynamic
instability of disks  is the
Papaloizou-Pringle  instability
(\cite{pp84};
\cite{pp85};  hereafter PP) which has
been extensively  studied both in
linear  theory  and by numerical
simulations (\cite{b85};
\cite{d85}; \cite{fr88}; \cite{g88};
\cite{ggn86}; \cite{k87}; \cite{nar87}; 
\cite{k89};
 \cite{h91}; \cite{z86}; see review by
\cite{ng89}).
    In these studies,  the ``disk''
is taken to be a thin torus (finite
height)  or annulus (infinite along
$z$) with the radial width  much less
than the radius.
     The  corotation  radius, where
the phase velocity of the wave
$\omega_r/m$ equals the angular
velocity of the matter $\Omega(r)$,
has a crucial role in that a wave
propagating radially across it can be
amplified.
   However, significant growth of a
wave typically requires many passages
through the corotation radius and
this requires
{\it reflecting} inner and/or 
outer boundaries of
the disk.
   Thus the PP instability  depends
on inner and outer disk boundary
conditions which are artificial for
accretion disks.

   Recently, we  pointed out a new
Rossby  wave instability of
non-magnetized accretion disks
(\cite{pap1};  hereafter Paper I).
      Paper I shows that the local
WKB dispersion relation for the
unstable modes is closely analogous to
that for Rossby waves in planetary
atmospheres.
   Rossby vortices associated
with the waves are well known
in planetary atmospheres and
give rise for example to the
Giant Red Spot on Jupiter
(Sommeria, Meyers, \& Swinney 1988;
Marcus 1989, 1990).

    In the cases considered in Paper I,
the instability occurs when there is
a  ``bump'' in the radial variation
of ${\cal L}(r) \equiv (\Sigma
\Omega/\kappa^2) S^{2/\Gamma}$, where
$\Sigma(r)$ is the surface mass
density of the disk,
$\Omega(r)$ the angular rotation
rate, $S(r)$ the specific entropy,
$\Gamma$ the adiabatic index, and
$\kappa(r)$ the radial epicyclic
frequency.
    Such a bump may arise from the
nearly inviscid accretion of matter
with finite specific  angular momentum
$\ell$ onto a compact star or black
hole.
    The accreting matter tends to
``pile up'' at the centrifugal radius
$r_c = \ell^2/(GM)$ (with $M$ the
mass of the central object) where it
gives a radially localized bump in
${\cal L}(r)$.

    In accretion
disks in some systems,
such as active galactic nuclei,
 the
weak self-gravity at
large distances can be important
in forming narrow rings (Shlosman, Begelman, \&
Frank 1990).
   Such rings would give a bump
in ${\cal L}(r)$ of the form
we have assumed.
   The bump in ${\cal L}(r)$ is
crucial because it leads to trapping
of the wave modes in a finite range
of radii encompassing the corotation
radius.
    Thus,  reflecting inner and outer
disk radii required for the PP
instability are {\it not} required
for the  Rossby wave instability.
   In  contrast with studies of the
PP instability, we allow a  general
equation of state where the entropy
of the disk matter can vary with
radius.

    The present paper develops in greater
detail the work of Paper I and
presents the full linear theory
analysis of the  instability without
some simplifying assumptions made
there.
The physical nature of the Rossby 
wave instability
is shown in Figure \ref{fig:rossby} where
we have overlaid the velocity field 
onto the surface density variations.
     The basic theory is given in
\S \ref{sec:diskeq} and \S \ref{sec:diskperb},
and results are given in \S \ref{sec:results}.
   In \S \ref{sec:discuss} we discuss the details of
this instability and compare it with
other hydrodynamic instabilities.
   Conclusions of this work are
summarized in \S \ref{sec:conclu}.

\begin{figure*}[t]
\centering
\epsfig{file=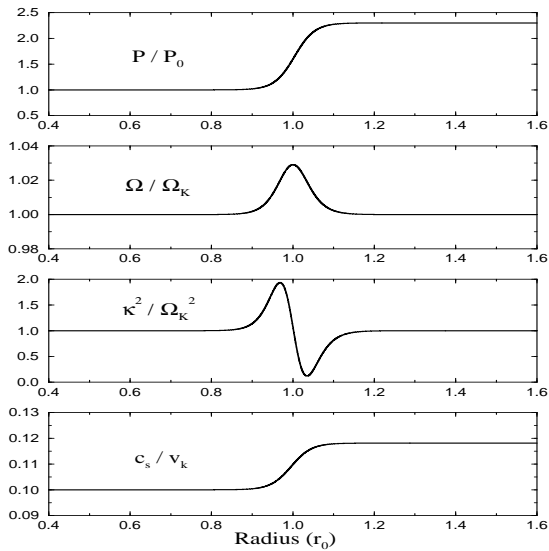,width=3in,height=3in}
\caption{Parameters of the initial
equilibrium for the homentropic step
jump case (`HSJ'): The pressure
$P(r)/P_0$; the corresponding angular
velocity
$\Omega(r)/\Omega_K$; the square of
the epicyclic frequency
$\kappa^2(r)/\Omega_K^2$; and the
effective sound speed $c_s/v_K$.
The pressure  jump is derived from a jump
in surface density $\Sigma(r)$ with
an amplitude
${\cal A} = 0.65$ over $\Delta r/r_0
= 0.05$. This results in a pressure
jump $(1+{\cal A})^\Gamma
\approx 2.3$ for
$\Gamma = 5/3$.
   The disk rotation is close to
Keplerian except near the region of the
pressure jump.
  Note that $\kappa^2$
is everywhere positive.
\label{fig:hsj-equil}
}
\end{figure*}

\section{Equilibrium Disks}
\label{sec:diskeq}

    We consider the stability of
non-magnetized accretion disks.
   The disks are assumed geometrically
thin with vertical half-thickness
$h \ll r$, where $r$ is the radial
distance.
    We use an inertial cylindrical
$(r,\phi,z)$ coordinate system.
    The surface mass density and
vertically integrated pressure are:
$$\Sigma(r) = \int_{-h}^{h} dz
~\rho(r,z)~,
\quad \quad P(r)= \int_{-h}^{h} dz
~p(r,z)~.$$
       The equilibrium disk is
stationary ($\partial/\partial t=0$)
and axisymmetric ($\partial/\partial
\phi = 0$), with the flow velocity
${\bf v} \approx v_\phi(r)
\hat{\phi}$; that is, the accretion
velocity $v_r$  is
assumed negligible.
    Self-gravity of the disk
is assumed negligible so that
$\Phi(r) = - GM/(r^2 + z^2)^{1/2}$,
where $M$ is the mass of the central
object. But we will discuss the
effects of self-gravity in \S \ref{sec:discuss}.

For the
axisymmetric
equilibrium disk, the radial
force balance is:
\begin{equation}
\label{eq:r-balance}
\frac{v_\phi^2}{r} \equiv r \Omega^2
=
\frac{1}{\Sigma} \frac{dP}{dr}  +
\frac{d\Phi}{dr}~~,
\end{equation}
The vertical hydrostatic
equilibrium gives
$h(r) \approx (c_s/v_\phi)r$, where
\begin{equation} c_s^2  \equiv \Gamma
\frac{P}{\Sigma}
\end{equation}
the square of adiabatic  sound speed
and $\Gamma$ is an effective
adiabatic index  discussed
further in \S \ref{sec:diskperb}.

   The  focus of this paper is on the
stability of equilibrium disks
with a slowly varying
background shear flow {\em and} a
finite-amplitude density variation over a
finite radial extent.
    Specifically,
we envision conditions where
inflowing matter
accumulates at some radius, for example, a
centrifugal barrier.
  This accumulation may occur if
matter is supplied at a rate
exceeding the rate at which
matter spreads by say a turbulent
viscosity.
   Depending
on the  angular momentum
distribution of the incoming flow,
newly supplied matter can  form a
disk with a finite ``bump''
or ``jump'' in the surface density at say $r$
with a radial width $\Delta r$.
    When the disk  surface
density is sufficiently small,
the disk optical depth is small
compared with unity and cooling
by radiation is efficient.
   As more  matter is
accumulated, the disk
becomes optically thick so that
heat builds up inside the disk 
and can be vertically
confined, then
pressure forces start to
have an important  role for the
disk dynamics.
   The present work is directed
at  optically thick disks  with
significant pressure forces.

   We model the mentioned ``bumps''
and ``jumps''
with fairly general
radial profiles of
$\Sigma(r)$ and $P(r)$.
   These profiles are used to
obtain the corresponding  $\Omega(r)$
profile using equation
(\ref{eq:r-balance}).
   Specifically,
we consider two
surface density distributions:
One is has a ``step jump''
from
$\Sigma_1$ to $\Sigma_2 ~>~
\Sigma_1$  over $\Delta r$ and this
jump is surrounded by a smooth
background flow,
\begin{equation}
\label{eq:sj-rho}
{\Sigma \over \Sigma_{\ast}} =
1 + \frac{\cal A}{2}
\left[\tanh\left(\frac{r-r_0}
{\Delta r}\right) + 1 \right]~~,
\end{equation}
where $\Sigma_\ast =
\Sigma_0 (r/r_0)^{-\beta}$ is the
surface density for the background disk,
$\Sigma_0$ is its value at
$r_0$ and $\beta$ characterizes
the slope (it is $-3/4$ in
a standard $\alpha$ disk model).
Quantities ${\cal A}$ and
$\Delta r$  measure
the amplitude and width of the jump
respectively, and $r_0$ is radius
of the  jump.
The second case we consider is a
``Gaussian bump,''
\begin{equation}
\label{eq:gau-rho}
{\Sigma \over \Sigma_\ast} = 1 + ({\cal A}-1)
\exp\left[-{1\over
2}\left(\frac{r-r_0} {\Delta
r}\right)^2\right]  ~,
\end{equation}
where again, ${\cal
A}$ and $\Delta r$ measure the height
and width of the bump, respectively.

   We assume an ideal gas equation of
state,  $P \propto
\Sigma T$.
   Depending on whether the
flow is homentropic\footnote{In actual
fluids, the pressure   depends
on say both density $\Sigma$  and entropy $S$.
   In this
paper, we use the term {\em homentropic} to
indicate that the pressure depends
only on density with the entropy
 a constant, $P=P(\Sigma)$.
   The
term {\em barotropic} is sometimes
used for the same situation.
    Similarly, we use {\em
nonhomentropic} (instead of {\em
nonbarotropic}) to describe a flow
where the entropy is not a constant
and $P=P(\Sigma, S)$. }
or not, we
can further derive or specify the
radial distributions of temperature
$T(r)/T_0$ and pressure $P(r)/P_0$ based on
$\Sigma(r)$, where $T_0$ and $P_0$
are the values at
$r_0$ for the background disk.
   We also
set $v_0 \equiv r_0 \Omega_0$ to be
the  Keplerian speed at $r_0$, and
define the dimensionless  sound speed
as $c_0^2 \equiv
\Gamma (P_0/\Sigma_0)/ v_0^2$. In
most of the following analysis, we
take
$r_0 = 1$, $\Delta r/r_0 = 0.05$,
$\Gamma = 5/3$, and $c_0 /v_0= 0.1$.

\begin{figure*}[t]
\centering
\epsfig{file=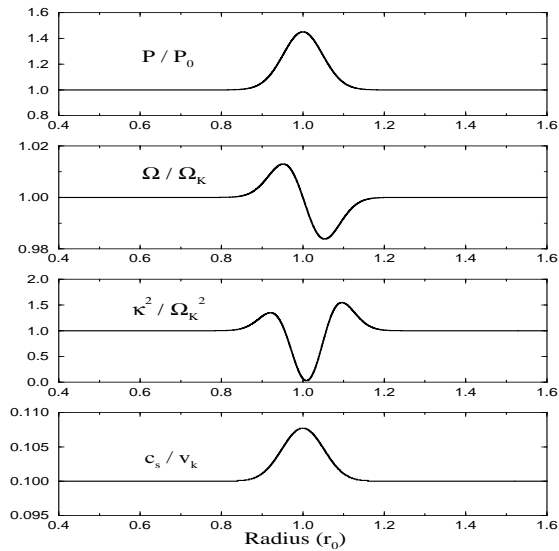,height=3in,width=3in,angle=0}
\caption{
Similar to Figure
\ref{fig:hsj-equil} but for the
case of an
homentropic Gaussian bump
(`HGB').
    The peak of the surface
density  is ${\cal A} = 1.25$
at $r=r_0$,
which gives a peak of
the pressure  ${\cal
A}^\Gamma
\approx 1.45$ for $\Gamma = 5/3$.
Again,
$\kappa^2$ is everywhere positive.
\label{fig:hgb-equil}
}
\end{figure*}

\begin{figure*}[t]
\centering
\epsfig{file=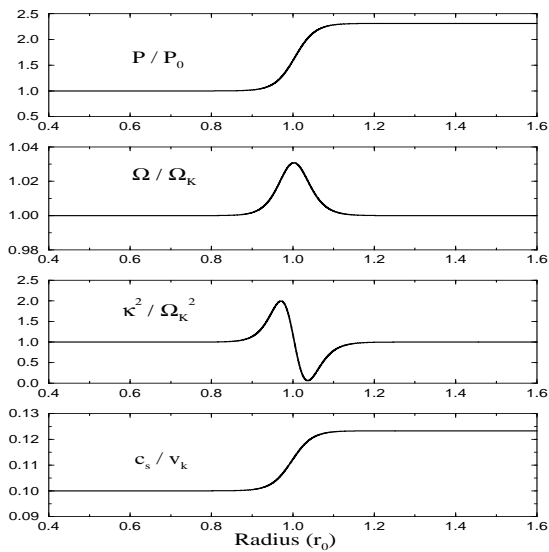,height=3in,width=3in,angle=0}
\caption{
Similar to Figure
\ref{fig:hsj-equil} but for the
case of a
nonhomentropic step jump
(`NSJ'). The jump amplitude of the
surface density and temperature is
${\cal A} = 0.52$, which gives a
pressure jump of
$(1+{\cal A})^2 \approx 2.3$. Again,
$\kappa^2$ is everywhere positive.
\label{fig:nsj-equil}
}
\end{figure*}

\subsection{Homentropic Disks}

For homentropic flows, we can
determine
$P/P_0$ and $T/T_0$ using
$\Sigma/\Sigma_0$ alone because of
the
relations $P/P_0 =
(\Sigma/\Sigma_0)^{\Gamma}$ and
$T/T_0 = (P/P_0)/(\Sigma/\Sigma_0)$.
   Based on these, we present two sample
initial configurations: one is a step
jump with ${\cal A} = 0.65$ which we
call case `HSJ' (homentropic step
jump);
the other is a Gaussian bump
with ${\cal A} = 1.25$ which we call
case `HGB'  (homentropic Gaussian
bump).
  The adiabatic index is
$\gamma = 5/3$ and the width $\Delta
r/ r_0 = 0.05$.
Figures
\ref{fig:hsj-equil} and
\ref{fig:hgb-equil} show the profiles
of $P/P(r_0)$,
$\Omega(r)/\Omega_K(r)$,
$\kappa^2(r)/\Omega_K^2(r)$,
and $c_s/v_K(r_0)$ for
the `HSJ' and `HGB' cases.

Here, we have chosen the 
parameter $\beta$
to be zero in the smooth 
profile of $\Sigma_\ast$.
We have studied the 
dependence of our results
on $\beta$ and find that 
for $ 0 \leq \beta \leq 3/4$,
our results are essentially
independent of $\beta$.
Thus, we omit further discussion
of the background
disk.

\subsection{Non-Homentropic Disks}

   For nonhomentropic flows, we have to
specify $T/T_0$ in addition to
$\Sigma/\Sigma_0$ in order to
determine
$P/P_0$ and consequently $\Omega(r)$.
   Such conditions were studied in Paper I
for a Gaussian bump, which we will
not repeat here.
  Instead, we
study a case where there a
simultaneous step jump in both
surface density and temperature.
   In
order to avoid too many parameters,
we will assume that the width and
amplitude of the jumps are the same
for surface density and temperature,
which are described by equation
(\ref{eq:sj-rho}).
   We will call this
initial configuration as case `NSJ'
(nonhomentropic step jump).
  Figure
\ref{fig:nsj-equil}  shows the
dependences of different
variables with ${\cal A} = 0.52$ and
$\Delta r/ r_0 = 0.05$.

As we  show later, the nonhomentropic
cases have some quantitative
differences from the homentropic
cases but the essential physics is
the same.

\section{Perturbations of Disk}
\label{sec:diskperb}

    We consider small perturbations
to the inviscid Euler equations.
    The perturbations  are considered
to be in the
plane of the disk.
  Thus  the
perturbed surface mass density is:
$\tilde{\Sigma} = \Sigma +
\delta \Sigma(r,\phi,t)$; the
perturbed vertically integrated
pressure is:
$\tilde{P} = P+\delta P(r,\phi,t)$;
and the perturbed flow velocity is:
$\tilde{\bf v} = {\bf v} +\delta {\bf
v}(r,\phi,t)$, with
${\bf \delta v} = (\delta v_r,\delta
v_\phi,0)$.
   The equations for the 2D
compressible  disk are:
\begin{equation}
\label{eq:mascon} {D
\tilde{\Sigma}\over Dt} +
\tilde{\Sigma}~ {\bf \nabla}\cdot
\tilde{\bf v} = 0~,
\end{equation}
\begin{equation}
\label{eq:momcon} {D \tilde{\bf
v}\over Dt}  = -{1\over
\tilde{\Sigma}} {\bf \nabla}
\tilde{P} - {\bf
\nabla}\Phi ~,
\end{equation}
\begin{equation}
\label{eq:enpy} {D \over Dt} \bigg
[{\tilde{P} \over
\tilde{\Sigma}^\Gamma}\bigg] = 0~,
\end{equation}
where
$${D \over Dt} \equiv {\partial \over
\partial t} + {\bf v}\cdot {\bf
\nabla}$$
(see Paper I). Here, $S \equiv
P/\Sigma^\Gamma$ is referred to as
the entropy of the disk matter.
   Equation (\ref{eq:enpy})
corresponds  to the adiabatic motion
of the disk matter.
   The rigorous relation for the
perturbation of the disk is of course
$D(p/\rho^{\gamma_t})/Dt = 0$, where
$p, \rho$ and $\gamma_t$ are the 3D
pressure, density and adiabatic
index, respectively. (We reserve the
symbol
$\gamma$ for mode growth rate.)
  For slow perturbations of the disk
matter (time scales $< h/c_s \sim
1/\Omega$), we have $h \propto
\sqrt{T}$ so that
$T \propto
\Sigma^{2(\gamma_t-1)/(\gamma_t+1)}$,
and consequently $P \propto
\Sigma^\Gamma$ with
$\Gamma = (3\gamma_t-1)/(\gamma_t+1)$.
   Equations (\ref{eq:mascon}) -
(\ref{eq:enpy}) are the  vertically
integrated 2D Euler equations.
   The equations assume that  the
disk thickness does not change
rapidly with radius or time, $|dh/dr|
\ll 1$ and $|\partial h /\partial t|
\ll \Omega r$.
   The above equations are closed
with an equation of state.
   The ideal gas law is applicable
for the conditions of interest
$P = \Sigma T/\mu$, where $\mu$ is
the mean mass per particle.
   Notice that  the
pressure is in general
a function of  {\it
both} the surface density $\Sigma$
and the entropy $S$ (or
equivalently, the temperature $T$).
   This is  general in contrast with the
commonly made assumption  that
$P=P(\Sigma)$.

    Following the steps of Paper I,
we linearize the equations by
considering  perturbations $\propto
f(r)\exp (im\phi - i \omega t)$,
where
$m=\pm1,\pm2,$ etc. is the azimuthal
mode number and
$\omega=\omega_r + i \gamma$  is the
mode  frequency.
  We  use
$\Psi \equiv \delta P/\Sigma$ as our
key variable which is analogous  to
enthalpy of the flow.
  (In a homentropic flow
considered by PP, $\Psi$ is the enthalpy.)
    The basic equations are given in
Paper I, but for
completeness  we also give the main
equations here:
\begin{equation} i\Delta \omega
\delta \Sigma ={\bf
\nabla}\cdot (\Sigma \delta {\bf v})~,
\end{equation}
which is from the continuity equation
(\ref{eq:mascon});
\begin{equation}
\label{eq:rho}
\delta \Sigma = \frac{\Sigma}{c_s^2}
\Psi  + i~\frac{\Sigma \delta v_r}
{\Delta \omega L_s}~,
\end{equation}
\begin{equation}
\label{eq:vr}
\Sigma \delta v_r = i~{\cal F}
\left[{\Delta \omega \over
\Omega}\left(\Psi^\prime- {\Psi \over
L_s}\right)
 -2k_\phi \Psi \right]~,
\end{equation}
\begin{equation}
\label{eq:vp}
\Sigma \delta v_\phi \!=\! {\cal F}
\bigg[ -k_\phi\left({\Delta \omega
\over \Omega}  + {c_s^2
\over
\Delta \omega \Omega
L_sL_p}\right)\Psi  +{\kappa^2\over 2
\Omega^2}\left(\Psi^\prime-
{\Psi\over L_s}\right) \bigg],
\end{equation}
which are from Euler's equation
(\ref{eq:momcon}).
Here,
\begin{eqnarray} k_\phi & \equiv & {m
\over r } ~,
\nonumber \\
\kappa^2 & \equiv & {1 \over r^{3}
}{d(r^4
\Omega^2)\over dr}~,
\nonumber \\
\Delta \omega(r) &\equiv& \omega -
m\Omega(r) =
\omega_r-m\Omega(r)+i\gamma ~,
\nonumber \\ {\cal F}(r) & \equiv &
\Sigma ~\Omega
\big/ [\kappa^2 - \Delta \omega^2
-c_s^2 /( L_sL_p)]~,
\nonumber \\ L_\Sigma & \equiv & 1
\bigg/\bigg[{d
\over dr}~ {\rm ln}\Sigma \bigg] ~,
\nonumber\\ L_s & \equiv & \Gamma
\bigg/
\bigg[{d \over dr}~{\rm ln}
\bigg( {P\over \Sigma^\Gamma }\bigg)
\bigg]~,
\nonumber \\ L_p & \equiv & \Gamma
\bigg/\left[ {d
\over dr}~ {\rm ln}(P)\right]~~,
\end{eqnarray}
where
$|L_\Sigma|$, $|L_s|$ and $|L_p|$ are
the radial length scales of the
surface density, entropy  and
pressure variations, respectively.
They are related as
\begin{equation}
\frac{1}{L_p} = \frac{1}{L_s} +
\frac{1}{L_\Sigma}~~.
\end{equation}

Furthermore, we can obtain
$$ {1\over
r}\left({r {\cal F} \over
\Omega}
\Psi^\prime\right)^\prime
-{k_\phi^2{\cal F}\over
\Omega} \Psi ={\Sigma \Psi \over
c_s^2} + {2k_\phi {\cal
F}^\prime\over
\Delta \omega}\Psi$$
\begin{equation}
 +\left[{{\cal F} \over
\Omega L_s^2}+ {1\over
r}\left({r{\cal F}\over \Omega
L_s}\right)^\prime +{ 4 k_\phi {\cal
F} \over \Delta \omega L_s}+
{k_\phi^2 c_s^2 {\cal F} \over
\Delta \omega^2 \Omega L_sL_p}
\right]\Psi ~~,
\end{equation}
(see Paper I), which can in turn be
written as
\begin{equation}
\label{eq:master}
\Psi'' + B(r) \Psi^\prime + C(r) \Psi
= 0~~,
\end{equation} where
\begin{eqnarray} B(r) & = &
\frac{1}{r} +
\frac{{\cal F}'}{\cal F} -
\frac{\Omega'}{\Omega} \\ C(r) & = &
-c_1-c_2 ~~,
\end{eqnarray} and
\begin{eqnarray} c_1\! &=& \!\!\!k_\phi^2 +
\frac{\kappa^2 -
\Delta \omega^2}{c_s^2} + 2 k_\phi
\frac{\Omega}{\Delta \omega}
\frac{{\cal F}'}{\cal F} ~~,\\ 
c_2\! &=&\!\!\!
\frac{1-L_s'}{L_s^2}  + \frac{B(r)+4
k_\phi \Omega /
\Delta \omega}{L_s}  + \frac{k_\phi^2
c_s^2 / \Delta
\omega^2 - 1}{L_s L_p}~~.
\end{eqnarray}

   For homentropic flow,
$L_s \rightarrow
\infty$.
   Thus the coefficients in
the  above equations simplify to
give
\begin{eqnarray} {\cal F}(r) &= &
{\Sigma ~\Omega \over
\kappa^2 - \Delta \omega^2} ~~,\\
C(r) &=& - c_1 ~~.
\end{eqnarray}
In this limit, our
equation
(\ref{eq:master}) is the same as
that given previously
(for example, PP).

   For any given  equilibrium disk,
we can  answer the following
questions:  (1) Are there unstable
modes
 with positive growth
rates $\gamma > 0$?
(2) What is the nature of
the radial wave functions $\Psi(r)$?
(3) What is
the dependence  of $\omega_r$ and
$\gamma$ on the initial equilibrium?
(4) And, what is the physical
mechanism(s) of the instability.
    Equation (\ref{eq:master})
allows the determination of
 $\omega= \omega_r+ i\gamma$
and identification of mode structure
for  general disk flows, both stable
or unstable.

\begin{figure*}[t]
\centering
\epsfig{file=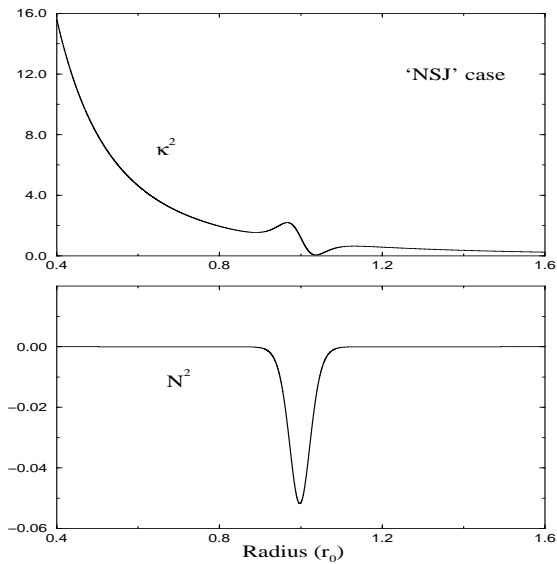,height=3in,width=3in,angle=0}
\caption{
A comparison of  $\kappa^2$
(upper panel) and
$N^2$ (lower panel) for the
nonhomentropic step jump ( `NSJ',
see Figure
\ref{fig:nsj-equil}).
  Note the
difference in magnitude of the two
frequencies.
  The disk
is locally stable according to the
Solberg-Hoiland criterion,
$\kappa^2 + N^2 > 0$.
\label{fig:k2n2}
}
\end{figure*}

\subsection{Axisymmetric
Stability}
\label{sec:rayleigh}

   It is important to know if the
equilibrium disk including
the ``jump'' or ``bump'' is stable to
axisymmetric perturbations.
  Rather than solving the non-local
axisymmetric stability problem (which
can be obtained from equation 15),
we consider the local
stability criterion for axisymmetric
perturbations. This condition
is expected
to be a sufficient condition for
non-local stability.
   If the disk pressure is
neglected, then the Rayleigh
criterion
$\kappa^2 \equiv (1/r^3)d(\Omega^2
r^4)/dr \geq 0$ implies  axisymmetric
stability  (Drazin \& Reid 1981, ch.
3;  Binney \& Tremaine 1987,
ch. 6).
    The more general condition
including the pressure is the
Solberg-Hoiland criterion,
\begin{equation}
\label{eq:k2n2}
\kappa^2(r)+N^2(r) \geq 0~,~ {\rm
where}~ N^2
\equiv
\frac{1}{\Sigma}\frac{dP}{dr}
\left(\frac{1}{\Sigma}
\frac{d\Sigma}{dr} - \frac{1}{\Gamma
P}
\frac{dP}{dr}\right)
\end{equation}
(Endal \& Sofia 1978).
Here, $N$ is the radial
Brunt-V$\ddot{a}$is$\ddot{a}$l$\ddot{a}$
frequency for the disk
 due to the radial entropy
variation.
  For homentropic flow,
$N^2$ is zero.
   Typically $N^2(r)$ is
negative for standard disk models
(Shakura \& Sunyaev 1973).
   Although
this is an unstable situation for
convection  in the
absence of rotation, it is stabilized
by the rotation since $|N^2|$ is
usually much smaller than
$\kappa^2$ in an approximate
Keplerian disk.
  Figure
\ref{fig:k2n2} shows illustrative
profiles of
$\kappa^2(r)$ and $N^2(r)$ for the
case 'NSJ' with
${\cal A}=0.52$.
   The quantity
$\kappa^2 + N^2$ remains positive.

As the amplitude of the bump/jump
${\cal A}$ increases, we can find a
critical value where condition  given
by equation (\ref{eq:k2n2}) is
violated.
  This happens at ${\cal
A}_{\rm crit}
\approx 0.74, 0.55, 1.26$ for  cases
'HSJ', 'NSJ' and 'HGB', respectively.
   Physically, this means that the local
specific angular momentum profile is
perturbed enough that it liberates
energy when locally exchanging two
fluids radially.
   Actual axisymmetric instability
is expected to occur only at values
of ${\cal A}$ larger than these
${\cal A}_c$ values.
    Note also that the
critical values are dependent upon
some other parameters, such as
$\Delta r$, $\Gamma$ and whether the
flow is homentropic or not.
    The
important point we note  here is
that we find non-axisymmetric
instability at values of ${\cal A}$
appreciable less than the values
${\cal A}_c$.

\subsection{Methods of Solving
Equation (\ref{eq:master})}
\label{sec:method}

Since the eigenfrequency
$\omega$ is in
general complex,  equation (16) is
 a second-order differential
equation with complex coefficients
which are functions of $r$.
   If
we discretize equation
(\ref{eq:master}) on a grid $i = 1,..,
N$ and combine it with appropriate
boundary conditions,  the problem
becomes that of finding the complex roots of
the determinant of an $N\times N$
tridiagonal matrix.
   Upon finding
the roots, we can then solve for the
corresponding complex wave
function $\Psi(r)$.

      We use a  Nyquist method
to find the $\omega$ eigenvalues.
  For this we
integrate along a closed path in
the complex $\omega-$plane and
thereby find
regions where roots (and poles)
reside.
     We then use a Newton
root-finder to locate the roots
accurately.
  Once a root is found for
a particular set of parameters, we
can find other roots by varying the
parameters slowly.
   There are two
apparent singularities in equation
(\ref{eq:master}),  one is at the
corotation resonance where $\Delta
\omega = 0$, which occurs
only on the real $\omega$
axis.
   The other is where $W
\equiv \kappa^2 - \Delta \omega^2 -
c_s^2/(L_s L_p)=0$, which is a
generalized form of the Lindblad
resonance condition for $L_s$ finite.
   Goldreich et al. (1986) showed that
the first singularity is a real singularity,
whereas the second is spurious and is
actually a regular point.
They
considered the case when $L_s
\rightarrow \infty$, but their
conclusion still holds for $L_s$
finite.
  Thus, extra caution is
needed in searching for roots in the
complex plane  with $\gamma > 0$,
because the condition
$W = 0$ can be satisfied both when
$\gamma=0$ and
$\Delta \omega_r^2 = \kappa^2 -
c_s^2/(L_s L_p)$, or when
$\Delta \omega_r = 0$ and $\gamma^2 =
-\kappa^2 + c_s^2/(L_s L_p)$.
   The
latter case is generally only possible
when
$\kappa^2 - c_s^2/(L_s L_p) < 0$,
which can be true if the pressure
bump/jump is strong enough.

\subsection{Boundary Conditions}

  The inner and outer boundary on $\Psi(r)$
in equation (16) are important.
  In most
previous studies where a torus is
considered, the boundary condition is
that the Lagrangian pressure
perturbation vanishes at the free
surfaces (both above and below the torus
and at its radial limits).
  In the present problem,
the ``bump'' or ``jump'' is
embedded in a background
disk so that  the appropriate
boundaries are
necessarily at a large distance
from the bump but still
in the disk.
    Inspection
 of the function $C$ in equation
(\ref{eq:master}) shows that
$r/r_0 < 0.8$ and $r/r_0 > 1.2$
correspond to large distances
from the bump.  In these outer
regions we can
use a WKB representation of $\Psi$
with $\Psi=f(r)
\exp\{is\}$.
We then have
\begin{equation} i s'' - s'^2 + i B
s' +  C = 0~~,
\end{equation}
where a prime denotes
a derivative with respect to $r$.
   Away from the bump, the
 $s''$ term is small and it can be dropped.
  Thus, we find  for $r \ll
r_0$,
\begin{equation} s' \approx
\frac{m\Omega}{c_s} +
i~\frac{1}{2}\left[
-\frac{\Omega'}{\Omega}-\frac{c_s'}{c_s}\right]
\approx \frac{m\Omega}{c_s} +
i~\frac{15}{16}\frac{1}{r}~~.
\end{equation}
  Thus, toward the inner
boundary,
$|\Psi| = |\exp\{i\int s'dr\}|
\propto r^{-15/16}$, increasing as
$r$ decreases.
   This is indeed the
dependence found for
the calculated wavefunctions
discussed in later sections.
Similar procedure can be
applied to the outer
boundary too.

The physical meaning of this
radiative boundary condition
can be understood by studying the
dispersion relation obtained from the
WKB approximation at the inner and
outer parts of the disk.
   Let $k_r^2
\approx s'^2
\approx {\cal R}e(C)$, we have
\begin{equation} (\omega_r -
m\Omega)^2 = \kappa^2 + k_r^2 c_s^2~~.
\end{equation} Since $\kappa^2
\approx \Omega^2$ at both $r_1$ and
$r_2$ away from the perturbed region,
the dispersion relation is then simply
\begin{equation}
\omega_r =  m\Omega \pm \sqrt{\Omega^2 +k_r^2
c_s^2 }~~.
\end{equation}
As it turns out that
the most unstable mode usually has
$\omega_r$ close to corotation at
$r_0$, i.e.,
$m \Omega(r_0)$, the above equation
implies that  we need to  choose
``--'' at $r_1$ since $\omega_r \ll
m\Omega(r_1)$ and similarly ``+'' at
outer boundary.
   Furthermore, the
group velocity
$v_g = \partial \omega/\partial k_r
\approx \mp c_s k_r/|k_r|$ at inner
and outer radius, respectively.
Since  physically we expect the wave
to propagate {\em away} from  the
central region, this requires that
$v_g < 0$ at $r_1$ and
$v_g > 0$ at $r_2$, which means $k_r
> 0$ at both
$r_1$ and $r_2$.

 The computational results presented
here we take the inner boundary
to be at $r_1=0.6r_0$  and the outer
boundary at $r_2=1.4r_0$, where
$r_0$ is the radius of the ``bump''
or ``jump.''
  We find that there is essentially no
dependence of our results on the values of $r_1$
and $r_2$.

We have  experimented with other
types of boundary conditions such as
vanishing pressure or vanishing
radial velocity.
   The overall properties of
the unstable modes show very little
change, but the relative phase between
the real and imaginary parts of $\Psi$
of course changes.

\subsection{Initial Value Problem}

A  different approach to
solving the linearized equations
(\ref{eq:mascon} - \ref{eq:enpy}) is
to treat them as an initial value
problem.  We let
\begin{equation}
\vec{u} =
\left( \begin{array}{c} u_1 \\ u_2 \\
u_3 \\ u_4
\end{array} \right) \equiv \left(
\begin{array}{c} \delta
\Sigma/\Sigma_0 \\
\delta v_r/v_0\\\delta v_\phi/v_0
\\\delta P/P_0
 \end{array}\right) ~~~~,
\end{equation}
\noindent where all variables have
the same meaning as in \S
\ref{sec:diskeq}.
Assuming that the
$\phi$ dependence of
$\vec{u}$ has the form
$\exp(im\phi)$, we can rewrite
equations (\ref{eq:mascon} -
\ref{eq:enpy})
as a system of first
order PDE in one dimension, i.e.,
\begin{equation}
\frac{\partial
\vec{u}}{\partial t}  + G(r):
\frac{\partial
\vec{u}}{\partial r}  +
H(r):\vec{u} = 0~~,
\end{equation}
\noindent where it is straightforward
to write down the elements of matrices
$G$ and $H$.
  It can be shown that
these equations are hyperbolic.
$G$ or
$H$ can be diagonalized  individually
but not simultaneously.

We have written a code to solve
equations (28)
using a 2nd order MacCormack
method.
   The idea is to determine the
growth rates of different
initial perturbations.
   Of course,  solutions of (28) will
eventually be dominated by the
{\em most} unstable mode.
  From this solution
we can get the growth rate,  the
real frequency, and the
radial wave function
for all 4 variables.

   However, we recommend the method
described in \S
\ref{sec:method} because it allows a
 comprehensive search for
unstable modes and the corresponding
eigenfunctions.
     The initial value code for
equation (28)
can then be used to verify the
behavior of these modes.
  We omit a detailed
comparison of the results obtained
from both types of codes but merely
report that we have found excellent
agreement between them on  growth
rates, mode frequencies, and radial
structure of eigenmodes.

\begin{figure*}[t]
\centering
\epsfig{file=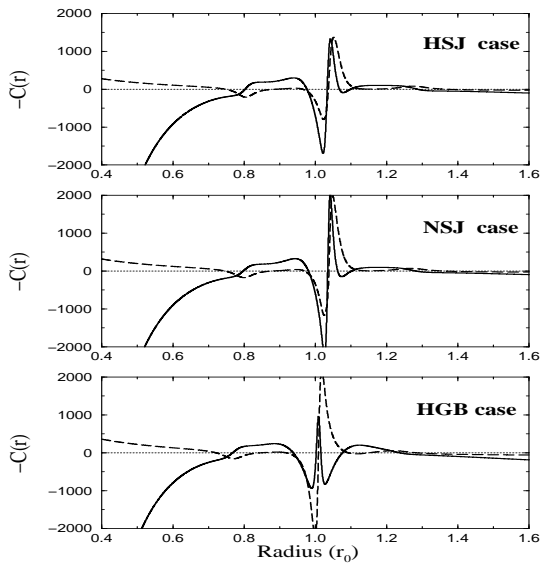,height=3in,width=3in,angle=0}
\caption{
The
effective ``potential well'' $-C(r)$ in
equation (\ref{eq:master}) for the
most unstable mode with $m = 3$,
found using parameters of cases
`HSJ', `NSJ', and `HGB'.
   The solid
and dashed curves are the real and
imaginary parts of the function $-C(r)$,
respectively.
   The unstable modes are
excited at radii near $r_0$ inside the
potential well.
  The unstable modes are trapped in
this potential well.
   This
trapping, however, is not absolute
since there is a finite ``probability''
for modes to tunnel through into both
the inner and outer parts of the disk.
\label{fig:potens}
}
\end{figure*}

\begin{figure*}[b]
\centering
\epsfig{file=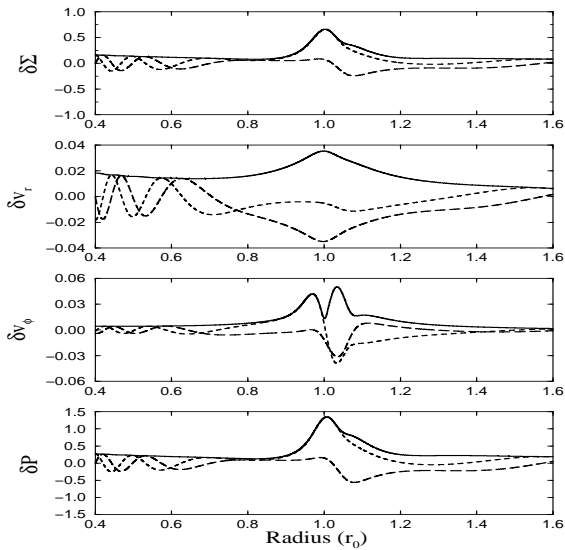,height=3in,width=3in,angle=0}
\caption{
Radial eigenfunction
for the perturbed surface density,
for the radial and azimuthal velocity
perturbations, and for the
pressure perturbation, for the
most unstable mode of the
homentropic step jump (`HSJ') for
$m=3$.
  The solid, dashed and
long-dashed curves  are the
amplitude, the real,  and the imaginary parts
of the eigenfunctions, respectively.
   All wave functions show their largest
amplitudes near the deepest position of
the potential well (slightly away
from $r_0$).
   The amplitudes
decrease going away from $r_0$.
The wave-like oscillations
towards the inner boundary is due to
the radiative
boundary
condition.
  Similar oscillations also
occur towards the outer
boundary, but their
amplitude is too small to be
evident.
   These features of the
wave functions
are  consistent
with the potential well shown in  the
upper panel of Figure
\ref{fig:potens}.
\label{fig:eigenfn-hsj}
}
\end{figure*}

\begin{figure*}[t]
\centering
\epsfig{file=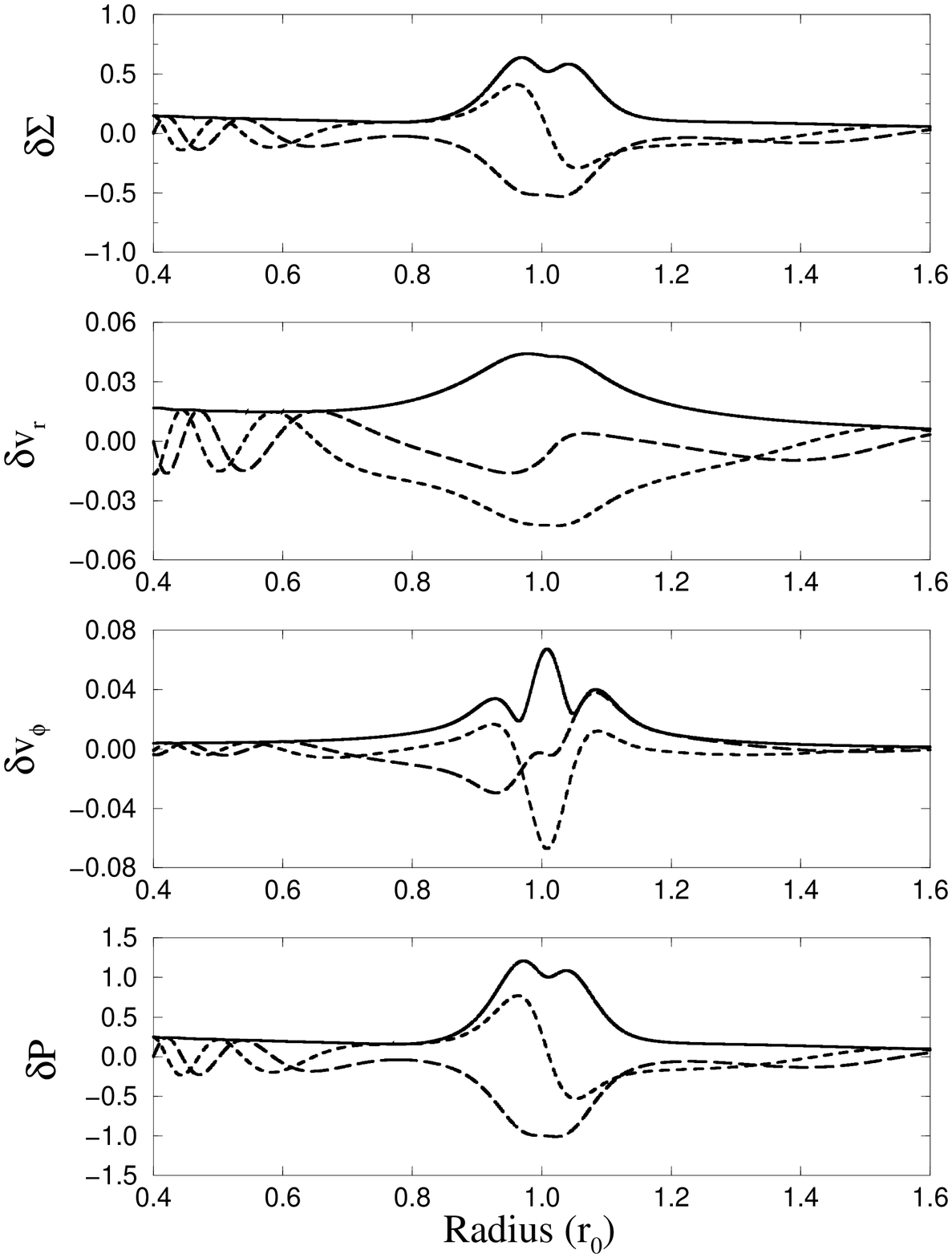,height=3in,width=3in,angle=0}
\caption{
Similar to Figure
\ref{fig:eigenfn-hsj} except that
parameters for the homentropic
Gaussian bump (`HGB')  are used.
   Again, the structure of the
eigenfunctions is consistent with
the potential well shown in the
bottom panel of Figure
\ref{fig:potens}.
   For example, the
two peaks   in the pressure
eigenfunction corresponds to the
two minima of the potential.
\label{fig:eigenfn-hgb}
}
\end{figure*}

\section{Results}
\label{sec:results}

We present the solutions to equation
(\ref{eq:master}) for different types
of initial equilibrium as discussed
in \S \ref{sec:diskeq}.
  In Paper I, we
showed that a necessary
condition for instability is that
the key function
${\cal L}(r) \equiv (\Sigma
\Omega/\kappa^2) S^{2/\Gamma}$
have an extreme as
a function of $r$.
  This
condition is a
generalized Rayleigh inflexion point
theorem for compressible and
nonhomentropic flow.
   All three
initial equilibria  presented
earlier satisfy this necessary
condition for instability.
Here we
discuss in detail the
behavior of these unstable
modes.

\subsection{Step Jump Case `HSJ'
with ${\cal A}=0.65$}

The initial equilibrium for this case
is shown in Figure
\ref{fig:hsj-equil}.
For $m=3$,  we find that
the most unstable mode
for this equilibrium  has
a growth rate
$\gamma/\Omega_0 \approx 0.154$
and
a real frequency
$\omega_r/(m\Omega_0) \approx 0.92$.

It is informative to plot the
effective ``potential,''
 $C(r)$ of equation
(\ref{eq:master}).  We
can neglect $B(r)$ because
its magnitude is
much smaller than $C(r)$.
  Thus,
equation (\ref{eq:master}) can be
simplified to give
\begin{equation}
\Psi'' + C(r) \Psi = 0~.
\end{equation}
 Then the function
$C(r)$ is analogous  to the
quantity $E-V(r)$
in quantum mechanics except that here
$C(r)$ is complex (Paper I).
  The real and
imaginary parts of $C(r)$
are shown in Figure
\ref{fig:potens} (the upper panel).
   First, we consider the region of
$0.95 < r/r_0 < 1.05$, which is most
affected by the presence of the
surface density jump.
   There is a
negative real part of $-C$ which is
analogous to the ``potential well''
in quantum mechanics where bound
states are possible, though again the
fact that $C$ is complex prevents a
direct analogy.
    The two sharp
``peaks'' in this region are not
singularities but result from the
two extrema of $1/(\kappa^2 - \Delta
\omega^2)$.
  In the region  $r/r_0 <
0.8$, the function $-C$ is dominated by
$c_1 \approx -(m\Omega/c_s)^2 \propto
- r^{-3}$,  whereas for $r/r_0 >
1.2$, $-C$ is  dominated by
$c_1 \approx  - (\omega_r/c_s)^2
\propto -$ const.
   Now consider an
unstable mode that is
excited in the ``potential
well'' around $r_0$.
   The positive
potential around
$r/r_0 = 0.9$ and $1.1$ causes this
mode to be evanescent in this region.
   The
potential is negative for
$r/r_0 < 0.8$ and $r/r_0> 1.2$ so
that
there will  be a finite probability for
this mode to ``tunnel'' through the
potential ``barriers''.
  In other
words, a mode excited around $r_0$
by the surface density
jump can ``radiate away'' into
both the inner and outer parts of the
disk.
   This is  indeed the
case as seen in
 Figure \ref{fig:eigenfn-hsj}
which shows the radial eigenfunctions of
various physical quantities derived
from $\Psi(r)$ using equations
(\ref{eq:rho})-(\ref{eq:vp}).
   In obtaining these eigenfunctions, we
have used outward propagating sound
wave boundary (i.e., radiative) 
conditions discussed
earlier.
   The relative phase shift
between real and imaginary parts
indicates this propagation.
   Note that the
``wavelength'' of the unstable modes
is at least $2 \Delta r \geq h$;
that is, the 2D approximation is
satisfied.
    Furthermore, we have maintained
the actual {\em relative} amplitude
among all 4 variables.

\begin{figure*}[t]
\centering
\epsfig{file=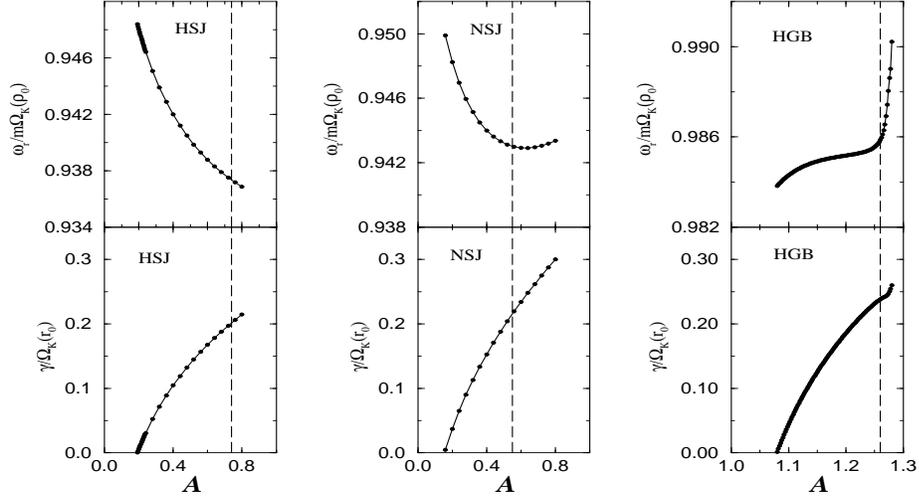,height=5in,width=3in,angle=-90}
\caption{
Dependences of the mode
frequencies and growth  rates on the
amplitude of the surface density
jump/bump ${\cal A}$ with $m=5$
for the three cases considered.
   The vanishing of the growth
rate for  ${\cal A}<
{\cal A}_{thres}$ indicates the
thresholds for individual cases.
   The
vertical dashed lines show the
critical values of ${\cal A}$ where
$\kappa^2 + N^2 = 0$.
    For larger values of ${\cal A}$
the flow violates the condition
for local
axisymmetric stability 
(see \S \ref{sec:rayleigh}).
\label{fig:modegamA}
}
\end{figure*}

\begin{figure*}[t]
\centering
\epsfig{file=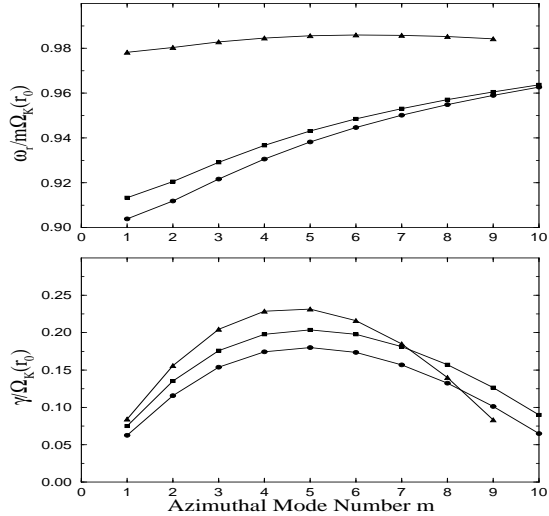,height=3in,width=3in,angle=0}
\caption{
Dependences of the mode
frequency $\omega_r$
and growth  rate $\gamma$ on the
azimuthal mode number $m$.
   The filled dots, squares
and triangles are for `HSJ', `NSJ'
and `HGB' cases, respectively.
\label{fig:modegamm}
}
\end{figure*}

\begin{figure*}[t]
\centering
\epsfig{file=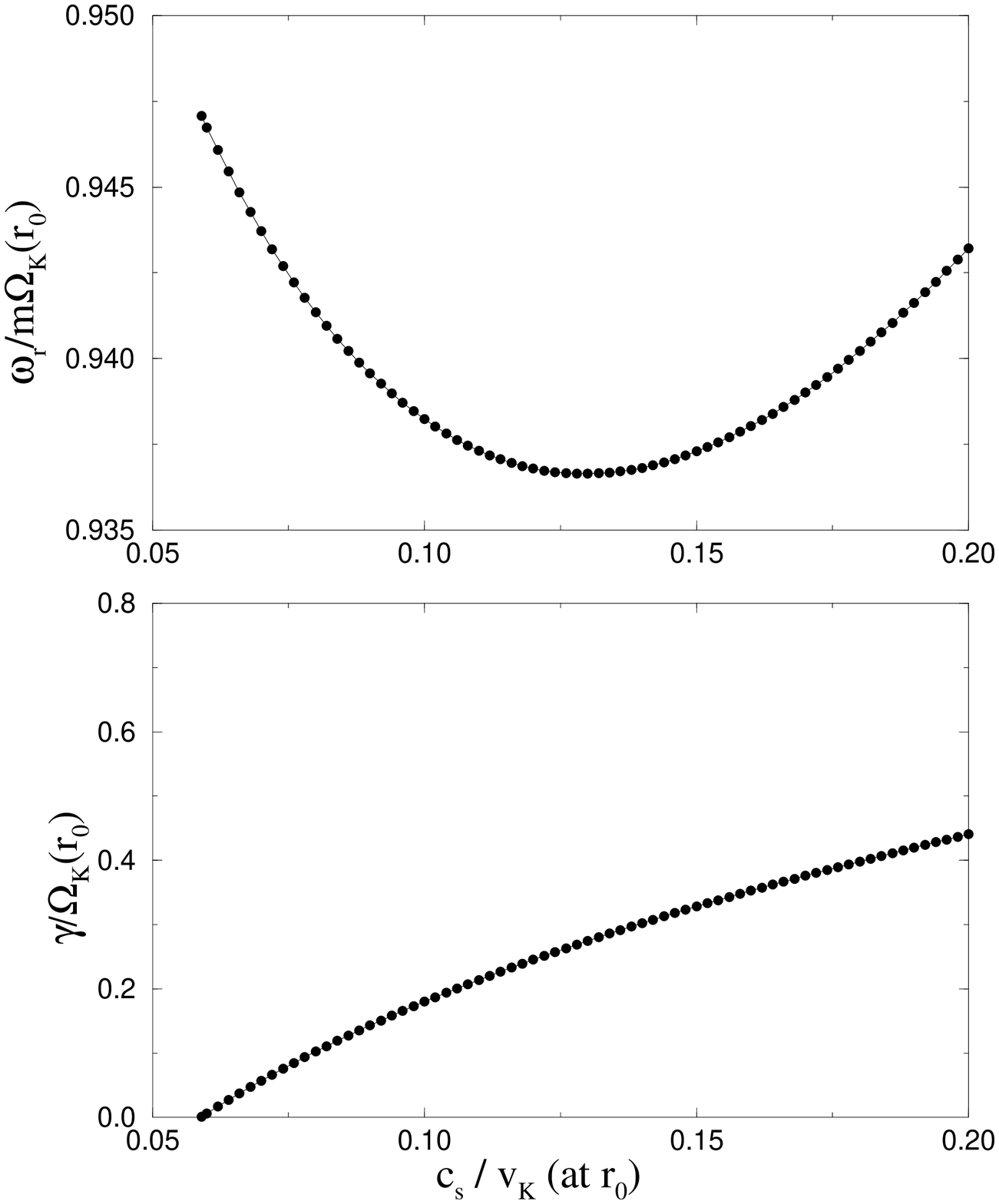,height=3in,width=3in,angle=0}
\caption{
Dependences of the mode
frequency $\omega_r$ and
growth  rate $\gamma$ on
 the sound speed for the
homentropic step jump
(`HSJ') for
$m = 5$.
\label{fig:gamma-cs}
}
\end{figure*}

\subsection{Step Jump Case `NSJ'
with ${\cal A} = 0.52$}

For $m=3$,  the most unstable mode
for this nonhomentropic equilibrium
has a growth rate
$\gamma/\Omega_0 \approx 0.176$,
and a real frequency
$\omega_r/(m\Omega_0)
\approx 0.93$.
   These values are slightly different
from the homentropic case presented
above with a slightly higher
growth rate.
   Note that the two cases
have roughly the same jump in total
pressure but the homentropic case
requires a higher surface density
jump. This will always be the case
if $\Gamma < 2$.
   Consequently, `NSJ' case has a
slightly deeper ``trap'' which is
shown in both the $\kappa^2$ and the
effective potential $-C$ as depicted
in Figure \ref{fig:potens} (the
middle panel).
   The eigenfunctions
for all 4 variables are almost the
same as in the case `HSJ' (cf. Figure
\ref{fig:eigenfn-hsj}),
therefore we do not
show them here.

   The comparison between cases `NSJ'
and `HSJ' confirms one of the
conclusions in Paper I that
the variation of
the temperature  is more
effective than that of the
surface density in driving this
instability.

\subsection{Gaussian Bump Case
`HGB' with ${\cal A}=1.25$}

For $m=3$,    the most unstable mode
for this case occurs with a growth rate
$\gamma/\Omega_0 \approx 0.21$,
and a real frequency
$\omega_r/(m\Omega_0) \approx 0.98$.
   The effective
potential $-C$ for this
case is shown in Figure
\ref{fig:potens} (the bottom panel).
The corresponding eigenfunctions are
shown in Figure
\ref{fig:eigenfn-hgb}.

    It is interesting to
note that for the
rather small Gaussian bump used here
(${\cal A}=1.25$),
the instability
has a higher growth rate than the
step jump cases considered above.
    This difference
can be understood as due to
the deeper
``trap'' produced by the Gaussian
bump.
   This can be seen by
comparing the $\kappa^2/\Omega^2_K$
profiles for both cases using Figure
\ref{fig:hsj-equil} and
\ref{fig:hgb-equil}.
   Since
$d(\Omega/\Omega_K)/dr$ is positive
at both the rising and declining
edges of a Gaussian bump, and is
positive only at the rising edge of a
step jump, the $\kappa^2$ profile has
two regions that are higher than
$\Omega^2_K$ in `HGB' compared to
just one such region in `HSJ'.
   In general, larger $\kappa^2$
indicates stronger stability,  and
they directly translate into the
``forbidden'' regions in the
potential structures shown in Figure
\ref{fig:potens} (the bottom panel).
     Thus, a mode excited around
$r_0$ is well trapped by the
``walls'' at $r/r_0 \approx 0.8-0.9$
and $1.1-1.2$.
   The `HSJ' case also
has a positive spike at $r/r_0
\sim 1.05$, but it does not provide
good trapping because the
spike is too narrow.
   Again, we have used the
outward propagating
sound wave boundary
conditions  discussed earlier
so that there is
a finite
probability for an unstable mode to
tunnel through the potential barriers.

\subsection{Instability Threshold and
Maximum Growth Rates}

  Consider now the dependence of the
growth rate and
mode frequency on the bump/jump
amplitude ${\cal A}$ and the azimuthal
mode number $m$.
    As the amplitude
${\cal A}$
decreases, the growth rate
of the instability is expected to
decrease.
   At small enough ${\cal A}>1$,
the instability should turn off (Paper I).
   It is clearly of
interest to determine
the minimum ${\cal A}$
for instability.

   Figure \ref{fig:modegamA} shows the
growth rate and
mode frequency as a
function of ${\cal A}$ for
the three cases  considered, all
for $m = 5$.
    The
threshold values are ${\cal A}_{\rm
thre} \approx 0.19, 0.16,$ and
$1.08$, for cases `HSJ', `NSJ', and
`HGB', respectively.
  Note that the
values of ${\cal A}_{\rm thre}$
depend on $\Delta r$ because
 the radial gradient of the
 surface density
$\Sigma(r)$  is important in
giving instability.
     Roughly speaking,
a factor of $\sim 1.2$
``jump''  or an $8\%$ ``bump''
in the surface
density are sufficient to cause
the Rossby wave instability.

   The vertical dashed lines in Figure
\ref{fig:modegamA} indicate the
${\cal A}$ values beyond which part
of the flow has $\kappa^2 < 0$ as
discussed in \S
\ref{sec:rayleigh}.
    In the step jump cases, the
the Rossby wave instability
 continues
smoothly through the point
where $\kappa^2$ changes sign.
   In case `HGB', the
instability also exists for values
of ${\cal A}$ where $\kappa^2<0$ in
part of the flow.
    There is a
continuous  increase in growth rate
as ${\cal A}$ increases (not shown
here), although there seems to be a
``kink'' at the point
where $kappa^2$ starts to have
a range of negative values.
    We
do not pursue here the situation
when both the Rossby
 and the axisymmetric
instabilities are present
simultaneously.
We emphasize that the Rossby
instability discussed here
has a substantial growth
rate ($\leq 0.2 \Omega_{\rm K}(r_0)$)
when the flow is stable
to axisymmetric perturbations.

Figure \ref{fig:modegamm} shows
the dependences of the growth
rate and
 mode frequency
on the azimuthal mode number
$m$ for the three cases.
   The peak of the growth rates
around $m = 4, 5$
probably results from a preferred
azimuthal wavelength in comparison
with the radial wavelength of the
unstable modes.

\subsection{Dependence on Sound Speed}

   Because the Rossby instability depends
critically on the pressure forces,
its growth rate has a strong
dependence on the magnitude of sound
speed.
   This is shown in Figure
\ref{fig:gamma-cs} for the `HSJ' case
for $m = 5$.
   As the sound speed
decreases, the pressure forces are no
longer strong enough to perturb the
rotational flow, the instability
disappears.
  This is similar to the
threshold seen in  Figure
\ref{fig:modegamA} when the pressure
jump/bump is too small.
   Judging from
the plot, the growth rate increases roughly
 linearly with $c_s$.

   Note that the  threshold value of
$c_s$ also depends on
$\Delta r$.
    As $c_s$ decreases,
so does the thickness of the disk.
   Thus, the allowed $\Delta r$ can be
smaller so long as the wavelength of
the unstable modes
is larger than the
disk thickness.  This
is a necessary condition for 2D
approximation to be satisfied.

\section{Discussion}
\label{sec:discuss}

\subsection{Origin of  Initial Equilibrium}

The condition for the Rossby wave
instability (hereafter, RWI) discussed
here can be understood in
terms of the Rayleigh's
inflexion-point theorem (Drazin \&
Reid 1981, p.81), which gives a
necessary condition for instability.
   If the accretion disk
is close to Keplerian
everywhere with temperature
and density being smooth powerlaws,
the RWI does
not occur.
   The local extreme value of
the key function
${\cal L}(r)$ considered here
is  caused by the
local step jump or Gaussian bump in
surface density (or possibly in
temperature if the flow  is
nonhomentropic).
   However, the considered
profiles of $\Sigma(r)$
and $P(r)$ are not the
only ones which may lead
to instability.
   For example, a
profile with a local
extreme in potential vorticity
distribution may also give
instability.
    An important aspect of
the RWI  is that the required
threshold for instability is quite
small.
   Typically, a $10-20\%$
variation of $\Sigma(r)$
over a lengthscale
slightly larger than
the thickness of the disk gives
instability.

The question is, then, can such an
initial equilibrium exist in real
astrophysical systems?
   The answer to
this question requires detailed
knowledge about how matter is
initially ``brought'' in towards the
gravitating object.
   One
plausible situation is that
accreting matter is gradually stored
at some large radial distance.
  Subsequently, after say a sufficient
build up of matter then
efficient accretion can proceed.
    This is perhaps the physical
situation
in the case of low mass X-ray
binaries where the systems go through
episodes of outbursts with long,
quiescent intervals between them (see
Tanaka \& Shibazaki 1996 for a
review).
   In protostellar disks, it is
possible that different regions of the
disk have different coupling strength
between matter and the magnetic field.
   This could  lead to accumulation of
matter at large distances
where the disk is non-magnetized.
  At smaller distances, the accretion
may be due to the Maxwell stress
from due to turbulent magnetic
fields arising from
the Balbus-Hawley
instability (Brandenburg 1998).
  Matter accumulation over some finite
extent in radius could most likely
lead to enhanced gradients of surface
density and/or temperature, thus
satisfying the  conditions for
the RWI discussed here.

Another aspect of the problem is the
role of vertical convection. We have
mostly discussed our instability in
the 2D limit where the vertical
variation of physical quantities is
all averaged away. In real disks,
however, there are certain processes
in the vertical direction that could
occur on a fast timescale, such as
vertical convection (\cite{paplin95};
\cite{khk99}). The vertical
convection is perhaps not a main
concern here since it helps to bring
the matter in a vertical column to
the same entropy, which is assumed in
our present study.  For the
horizontal motion, as we emphasized
before, the initial equilibria we
have studied are all stable to the
local convective  instabilities. So,
the RWI will
always be the dominant instability
unless there are other instabilities
(in the $\{r,\phi\}$ plane) that have
lower thresholds and faster growth
rates.

\subsection{Influence of Self-Gravity}

   For accretion disks in some
systems, for example, active galactic nuclei
and protostellar systems, self-gravity
may be important at large distances
and lead to axisymmetric instability.
  The WKB dispersion relation for
$m=0$ perturbations of a self-gravitating
disk is
\begin{equation}
\omega^2 = \kappa^2-2\pi G \Sigma |k_r|
+ k_r^2 c_s^2 ~
\end{equation}
(Safronov 1960; Toomre 1964).
   The least stable, smallest $\omega^2$
mode occurs for $k_r= \pi G\Sigma /c_s^2$,
and for this wavenumber, $\omega^2 = \kappa^2
(1-1/Q)$.
Thus, the disk is unstable if Toomre's
parameter $Q \equiv \kappa c_s/(\pi
G \Sigma) <1$.
  The threshold for instability $Q=1$
corresponds to a surface mass
density in the case of an AGN disk of
\begin{equation}
 \Sigma_{thres} = {\kappa c_s \over \pi G}
\approx 70.7 {\rm g \over cm^2}
\left({c_s/v_K \over 0.01}\right)
\left( {M \over 10^8 M_\odot} \right)^2
\left({1pc \over r} \right)^2~,
\end{equation}
where we assumed $\kappa \approx \Omega_K$
(Shlosman, Begelman, \& Frank 1990).

   Near the threshold with $Q$ somewhat less
than unity, this instability will act
to make axisymmetric rings
with radial wavelength
$\lambda_r \approx 2\pi c_s/\Omega_K =
2\pi h$.
   These rings have a form similar
to the case of our ``Gaussian bump.''
  Only a relatively small linear
growth of the rings is needed for
the non-axisymmetric RWI
to set in with a very
large growth rate.
   It is important to note that
the RWI is capable of transporting angular
momentum and thereby facilitating
accretion whereas the axisymmetric
instability  is not.

\subsection{Minimum Surface Density and
Relevance to  X-ray Novae}

We mentioned earlier the  important
role of the pressure
($\propto c_s^2$) for the RWI.
   This leads
to an important
physical requirement
on the minimum surface
mass density $\Sigma_c$
of the disk for the RWI to
be important.
   Heat must be confined
in the region of RWI
for a time much longer
than rotation period in order
that  the mode
grows without
 radiative
cooling.
   This confinement requires
a minimum surface density
which
depends on the opacity
and specific heat of matter.
    This translates to
roughly $\Sigma_c \sim
10^{2-3}$ g cm$^{-2}$.
   In other words, if $\Sigma$
is too small, the RWI will
not occur and there will be
no Rossby vortex induced
accretion.
   But once enough
matter has accumulated so that $\Sigma$
exceeds this threshold
$\Sigma_c$,  RWI sets in
and there is efficient accretion
due to Rossby vortices.
    This
may correspond to the
distinctly different states of
accretion in X-ray binary systems
where sources are often
observed to be ``active'' or
``quiescent''.

As shown by Tanaka \& Shibazaki (1996),
X-ray novae (especially
black hole candidates) show
long quiescent intervals in X-rays
despite that the companion star is still
continuously feeding mass to
the compact object at $\sim 10^{15-16}$
gm s$^{-1}$.
This strongly implies a mass
accumulation at some
large distances without much
accretion going on.
The accumulated
mass over an interval
of $\sim 50$ yrs will be
$\sim 1.5\times 10^{24-25}$ gm,
which gives a surface density of
$\Sigma \sim 5\times 10^{2-3}$
g cm$^{-2}$ with a size of
$\sim 3\times 10^{10}$ cm.
This critical $\Sigma$ is interestingly
close to the value we discussed above.
So, we believe it is worthwhile
to pursue RWI as an alternative
mechanism for causing an
outburst in X-ray novae, in addition
to the usual disk instability models.

\subsection{Comparison with the Related Work}

It is difficult to make a direct
comparison of the RWI with the original
Papaloizou
\& Pringle instability.
  In PP
instability, a torus and/or an
annulus  has two edges and  the
existence of certain unstable modes
critically depends  on these edges
(i.e., the so-called  ``principal
branch'').
   In our study, the surface
density bump case can perhaps be
viewed as a torus except that it is
embedded in a background Keplerian
shear  flow.
   Here, we discuss two
major differences between the two
instabilities.
   One is the treatment
of boundary conditions.
     In our study,
the propagating sound wave boundary
condition provides a natural and
important link of the unstable region
with the surrounding flow. This
allows us to directly apply such
instability to thin Keplerian
accretion disk of a much larger
radial extent.
   In the nonlinear
regime, this ``leakage'' will allow
the unstable modes to grow further
nonlinearly and impact a large part
of the disk flow also. The second
difference is the fact that, unlike
PP instability,  the rotational
profile $\Omega(r)$  is not taken as
a single power-law in our case.
   Instead, enhanced epicyclic frequency
occurs in association with the
surface density jump/bump boundaries.
This naturally creates the
``potential well'' that allows the
unstable modes to grow and provides
the trapping at the boundaries, as
illustrated by the structure of the
``potential'' in Figure
\ref{fig:potens}. The corotation
radius
occurs within this potential well.

\begin{figure*}[t]
\centering
\epsfig{file=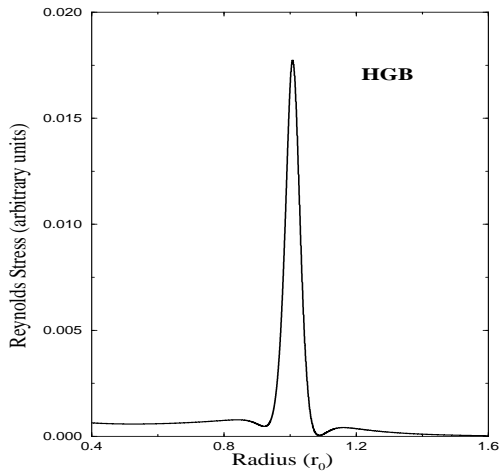,height=3in,width=3in,angle=0}
\caption{
The Reynolds stress
(arbitrary units) at radius $r$ as
derived from the velocity variations
in linear theory.
   The positive values of this
quantity indicates  outward
transport of angular momentum.
\label{fig:reys}
}
\end{figure*}

\subsection{Nonhomentropic vs. Homentropic}

In Paper I we have already given a
necessary condition for instability
with respect to 2D non-axisymmetric
disturbances, that is,  for a key
function
\begin{equation}
\label{eq:inst-cond} {\cal L}(r)
\equiv {\cal F}(r) S^{2/\Gamma}(r)
\end{equation} to have a local
extreme, where
${\cal F}^{-1} = \hat {\bf z}\cdot
({\bf \nabla}\times {\bf v}) /\Sigma$
is the potential vorticity, $S =
P/\Sigma^\Gamma$ is  the entropy.
This is a generalized Rayleigh
criterion which was originally
derived for incompressible flows (cf.
\cite{dr81}). Most of studies on PP
instability have assumed homentropic
flow where entropy of the whole fluid
system is a constant. The inclusion
of $S(r)$ in equation
(\ref{eq:inst-cond}) makes it perhaps
more applicable to real astrophysical
disks where entropy is usually not a
constant. As we have shown in this
paper,  the onset of the RWI,
however, does not depend
on having an entropy gradient.
   On the
other hand,  as shown in Paper I and
by the comparison of `HSJ' and `NSJ'
cases, temperature variation is more
effective in causing the instability.
Furthermore, entropy gradient
introduces more features into the
problem, most notably that the
potential vorticity of the flow is no
longer conserved due to the net
thermodynamic driving of vorticity
from the  fact that
${\bf \nabla}T \times {\bf \nabla}S
\ne 0$ (see Paper I).

\subsection{Implications for Angular
Momentum Transport}

   In order to evaluate the angular
momentum transport  from this
instability, it is necessary to
evaluate terms which include {\it
second} order terms.
   However, the
present linear theory is  valid only
to  first order.
    Instead of the angular momentum
transport, we
calculate the Reynolds
stress derived from the velocity
perturbations $<\delta v_r \delta
v_\phi>$.
  This term is expected to be
important for
the angular
momentum transport.
    The Reynolds stress at each radius
can be calculated as, by averaging
over $\phi$,
$$ {\cal R}_s(r)  = -
\int_0^{2\pi} d\phi~ r^2~
\Sigma_0~\frac{d\Omega}{dr}~{\cal
R}e(\delta v_r) {\cal R}e(v_{\phi})~,
$$
\begin{equation}
 =  - \pi r^2
\Sigma_0~\frac{d\Omega}{dr}~
\left[{\cal R}e(\delta v_r){\cal
R}e(\delta v_\phi)  + Im(\delta v_r)
Im(\delta v_\phi) \right]~~.
\end{equation}
The result is shown in
Figure
\ref{fig:reys} for the  `HGB' case
with $m = 5$  (results for `HSJ' and
`NSJ' cases are  similar).
    This
instability indeed causes a mainly
positive Reynolds stress which
corresponds to an outward
transport.

    A full evaluation
of the angular momentum and
matter  transport due to
the RWI evidently requires
non-linear hydrodynamic simulations.
    We have begun
to perform these simulations and will
present our findings in a forthcoming
paper (Li, Lovelace, \& Colgate
1999).
   Briefly, we have observed the
production of largescale 2D vortices
in the $\{r,\phi\}$ plane and these
vortices are observed to
survive the shear flow
for many  revolutions of
the disk (${\buildrel > \over \sim} 10$).
  There is
also indication of outward  angular
momentum transport, which shows
promises for this instability being a
robust mechanism of angular momentum
transport.

One important point we want to
emphasize is that the transport
process regulated by these large scale
vortices is inviscid, highly dynamic
and nonlocal.
   This is fundamentally
different from the standard
$\alpha$ disk models where
the flow is assumed to be
{\em viscous}, hence the transport is
at small scale (i.e., local) and a
quasi-stationary state can usually be
reached.
  In fact, the notion  of a
statistically stationary $\alpha$
value and  stationary accretion may
be an incorrect physical picture for
the accretion in some highly variable
systems such as X-ray binaries.

\section{Conclusions}
\label{sec:conclu}

   We have developed a
detailed linear
theory of the Rossby
wave  instability  associated
with an axisymmetric, local surface
density jump or bump in a thin
accretion disk.
    The instability is
termed Rossby
 due to its
WKB dispersion relation
which is analogous to that
for
Rossby waves in planetary
atmospheres.
   Rossby vortices associated
with the waves are well known
in planetary atmospheres and
give rise for example to the
Giant Red Spot on Jupiter
(Sommeria, Meyers, \& Swinney 1988;
Marcus 1989, 1990).
    The flow is made
unstable due to the existence of
local extreme value of a generalized
potential vorticity ${\cal L}(r)$
which includes
the radial variation of entropy.
   Depending on the parameters, the
unstable modes are found to have
substantial growth rates $\sim 0.2
\Omega(r_0)$, where $r_0$ is the
location of enhanced surface density
gradient.
   These modes are capable of
transporting angular momentum
outward.
   Since this instability
relies on  pressure forces
perturbing the rotational flow, it
 requires a minimum sound
speed, or a local ``heat'' content.
    We expect that the disk must be
optically thick in order for
the instability to be important.

   It is
important to understand the
nonlinear interactions between this
instability and the rest of the disk,
and to see whether the instability
is effective in
transporting angular momentum
globally.
   An important aspect of
the present instability
is that the unstable
modes propagate into the
surrounding stable disk flow,
thus allowing
the instability to impact a
large part of the disk.
    We have performed
preliminary nonlinear hydro
simulations  of this instability and
 will report the detailed results
in a forthcoming publication.
   The
simulations have confirmed  the
the linear growth of
the Rossby wave instability and they
have shown the vortices have
 long lifetimes (many orbital
periods).

\acknowledgements{ We acknowledge
useful conversations  with J. Frank,
S. Kato and C. Fryer. HL gratefully
acknowledges the support of an
Oppenheimer Fellowship.
   RL acknowledges support from
NASA grant NAG5 6311.
  This research
is supported by the Department of
Energy,  under contract
W-7405-ENG-36. }

\end{document}